\documentclass[aps,nopacs,onecolumn,preprintnumbers,amsmath,amssymb,longbibliography]{revtex4}

\usepackage{graphicx}
\usepackage{color}

\begin{document}

\title{Cooling of photoexcited carriers in graphene by internal and substrate phonons}
\author{Tony Low$^1$, Vasili Perebeinos$^1$, 
Raseong Kim$^2\footnote{Present address: Components Research, Intel Corporation, Hillsboro, OR 97124, USA}$, 
Marcus Freitag$^1$ and Phaedon Avouris$^1$}
\affiliation{$^1$IBM T.J. Watson Research Center, Yorktown Heights, NY 10598, USA\\
$^2$Network for Computational Nanotechnology, Purdue University, West Lafayette, IN 47907, USA}
\date{\today}

\begin{abstract}
{
We investigate the energy relaxation of hot carriers produced by photoexcitation
of graphene through coupling to both intrinsic and remote (substrate)
surface polar phonons using the Boltzmann equation approach.
We find that the energy relaxation of hot photocarriers in graphene on commonly used polar substrates,
under most conditions, is dominated by remote surface polar phonons.
We also calculate key characteristics of the energy relaxation process, such as the
transient cooling time and steady state carrier temperatures and
photocarriers densities, which determine the thermoelectric and photovoltaic photoresponse, respectively.
Substrate engineering can be a promising route to
efficient optoelectronic devices driven by hot carrier dynamics.
}
\end{abstract}

\maketitle

\section{Introduction}

Upon fast excitation of graphene carriers with light or other means, the dynamics of the resulting
non-equilibrium carrier distribution evolve on a fast time scale and has been extensively studied both experimentally\cite{SWDL08,MSCRS08,KPSFW05,BRE09,BKWM11,Newson_2009,Choi_APL_2009,Wang_APL_2008,George_NL_2008,Kumar_2009,Ishioka_2008,Seibert_1990,Winnerl_2011} and theoretically\cite{Winzer_2010,KPA11,Malic_PRB_2011}. The relaxation involves an initial fast evolution towards quasi-thermal distribution on a femtosecond timescale via electron-electron collisions\cite{Hertel_2000,Heinz_PRL_2010,Winzer_2010,KPA11}, followed by energy transfer to phonons on a longer picosecond timescale. 
The conversion of the excess energy of these photoexcited carriers
into electrical current before they lose this energy to the phonon baths represents one of the key challenges to
efficient optoelectronic device.  \\

In this paper, we study the energy relaxation pathways of the
photoexcited carriers via different inelastic scattering channels. 
Energy relaxation processes in graphene due to intrinsic optical and acoustic
phonons have already been studied\cite{BM09,TS09,Kubakaddi09,KPA11}. High energy optical phonon emission
by hot carriers is responsible for the subpicosecond fast cooling process\cite{SWDL08,MSCRS08,SAJ12},
followed by cooling via the acoustic modes. The latter is a slow process, that creates
an electron-phonon cooling bottleneck\cite{BM09}.
Here, we focus on an extrinsic mechanism for cooling of photoexcited carriers in graphene
via the remote surface polar phonon modes (SPP) of the substrate and compare their efficiency
under different conditions with those of the internal phonon modes.\\

In polar substrates such as SiO$_2$, a non-vanishing fluctuating electric field is generated by
the propagating surface phonon modes\cite{AM76}. 
The interactions of these SPP modes with charged carriers in the conduction channel
 was first explored in the context of
inversion layer of semiconductor-oxide interface\cite{WM72,HV79,FNC01}.
They have also been studied in other material systems such
as carbon nanotubes\cite{PR06,PRPA09,Bhupesh_2011},
where close proximity between charged carriers and
the underlying substrate renders the SPP-phonon scattering more prominent.
Similarly, in graphene, SPP was found to limit electronic transport properties\cite{MHY08,CJXIF08,FG08,PA10,KFJ10,Li_SPP,PHS12,SSG12,ZHKZ10,DZJ10}.
Recently, the SPP coupling with graphene plasmons
was also probed experimentally through infrared spectroscopy\cite{LW10,FAB11,CBGTH12}.  
In this work, we found irrespectively of the mechanism i.e. thermoelectric or photovoltaic,
that SPP limits the overall strength of the steady state photoresponse on common substrates,
and our results suggest that elimination of the SPP cooling channel can lead to an order 
of magnitude enhancement in the photoresponse.\\

In Sec.\,\ref{sec:theory}, we present the general theory,
where details of the models for the electron cooling power via the different phonon baths are presented in the Appendix.
We present the results of our calculation of the cooling powers
in Sec.\,\ref{sec:results}\,A and discuss their relative contribution in detail,
as a function of doping and electronic/lattice temperatures.
In Sec.\,\ref{sec:results}\,B, we apply the above models
to the study of the cooling dynamics of hot carriers
due to continuous or pulsed light excitations. We calculate
key experimental observables such as the transient cooling time,
and steady state quantities such as the non-equilibrium electronic
temperatures, excess photocarriers density and the out-of-plane thermal conductivity
for graphene on common substrates.

\vspace{2 mm}

\section{\label{sec:theory} Theory and Models}

Transition probability for emission and absorption of phonons
with a particular phonon bath $\alpha$ is described by the Fermi's golden rule, 
\begin{eqnarray}
S_{\alpha}(\bold{k},\bold{k}')=\frac{2\pi}{\hbar}\sum_{\bold{q}}\frac{1}{A}\left|M^{\alpha}_{\bold{k},\bold{k}'}\right|^2\left\{ N_{\omega_q}
\delta_{\bold{k}'-\bold{k}-\bold{q}}\delta(E_\bold{k}'-E_\bold{k}-\hbar\omega_{q})+
(N_{\omega_q}+1)
\delta_{\bold{k}'-\bold{k}+\bold{q}}\delta(E_\bold{k}'-E_\bold{k}+\hbar\omega_{q})
\right\}
\end{eqnarray}
where $\bold{q}$ is the phonon momentum,
$N_{\omega_q}=[\mbox{exp}(\hbar\omega_{q}/k_{B}T_L)-1]^{-1}$ is the Bose-Einstein distribution
and $M_{\bold{k},\bold{k}'}^{\alpha}$ are the transition matrix elements related to
the coupling with phonon bath $\alpha$, to be defined below.
For brevity, summation $\sum_{\bold{q}}\delta_{\bold{k}'-\bold{k}\pm\bold{q}}$ shall be implicit hereafter.
The cooling power is computed numerically by accounting for the 
transfer of electronic energy to the lattice during each 
scattering event i.e. $E_{\bold{k}}-E_{\bold{k}'}$. 
Therefore the net cooling power is calculated via\cite{conwell67,manion87,BM09,Kubakaddi09},
\begin{eqnarray}
\nonumber
{\cal P}^{\alpha}_{ss'} &=& \frac{g_s g_v}{A}\sum_{\bold{k}}\sum_{\bold{k}'} S_{\alpha}(\bold{k},\bold{k}')
(E_{\bold{k}}-E_{\bold{k}'})f_{\bold{k}}(1-f_{\bold{k}'})\\
\nonumber
&=& \frac{2\pi g_s g_v}{\hbar A^2} \sum_{\bold{k},\bold{k}'} \left|M^{\alpha}_{\bold{k},\bold{k}'}\right|^2
\delta(E_\bold{k}'-E_\bold{k}-\hbar\omega_{q})(E_{\bold{k}'}-E_{\bold{k}}) {\cal F}(k,k')\\
&=& \frac{g_s g_v }{(2\pi)^2\hbar } \int_{0}^{\infty}k dk \int_{0}^{\infty}k' dk' \int_{0}^{2\pi} d\theta
\left|M^{\alpha}_{\bold{k},\bold{k}'}\right|^2
\delta(k'-k-\omega_{q}v_{F}^{-1})(k'-k) {\cal F}(k,k')
\label{cpower}
\end{eqnarray}
where $T_{L}$ and $T_{E}$ are the lattice and electron temperatures respectively,
and the electron distribution function is described by $f_{\bold{k}}$, and we define
a composite Fermi-Boson distribution function as,
\begin{eqnarray}
{\cal F}(k,k')&\equiv& (N_{\omega_q}+1)f_{\bold{k}'}(1-f_{\bold{k}})-N_{\omega_q}f_{\bold{k}}(1-f_{\bold{k}'})
\label{comF}
\end{eqnarray}
As indicated by experiments\cite{SWDL08,MSCRS08,Heinz_PRL_2010}, the electronic system
is thermalized by the electron-electron interactions which occur at much
faster timescale than the electron-phonon processes we are calculating here.
Hence, it is appropriate to simply assume that
$f_{\bold{k}}$ follows the Fermi Dirac distribution function, i.e.
$[1+\mbox{exp}(\beta(E_{\bold{k}}-\mu))]^{\mbox{-}1}$ where $\beta=1/k_B T_E$
and $\mu$ is the chemical potential, controlled by chemical doping 
or electrical gating in experiments.
It is apparent that the composite electron-phonon distribution function
${\cal F}(k,k')$ becomes zero when $T_{E}=T_{L}$, hence zero cooling power.\\

In this work, we are interested in the energy exchange of electrons
with the different phonon baths i.e. intrinsic acoustic phonons (AP),
optical phonons (OP) and surface phonon polaritons (SPP).
Vibrations of the substrate ions with the opposite charge polarity
produce an electric field which decays exponentially away
from the surface. Carriers in the nearby graphene can feel this electric
field and be scattered by the SPP phonon. 
The decay length of the electric field is determined by the 
momentum transfer in the electron-phonon scattering event. 
For typical carrier density in graphene, the relevant momentum transfer
is of the order of nm$^{-1}$, such that a substantial coupling strength
is expected for graphene placed at van der Waals distance
of $3.4\,\AA$  away from the substrate. 
The transition matrix elements, $M_{\bold{k},\bold{k}'}^{\alpha}$,
for electron interaction with AP, OP and SPP phonons
are well-known in the literatures\cite{HS08,LPM05,ANDO06,FG08,FNC01,PA10}.
We therefore defer their discussions to the Appendix, with the
parameter set summarized in Table\,1. We shall
focus on the key results in what follows.

\begin{table}
\caption{Parameters for the optical (OP), acoustic (AP) and substrate phonons (SPP). 
For SPP, we consider SiO$_2$ and h-BN substrates. $\epsilon_{low}$ ($\epsilon_{high}$)
is the low (high) frequency dielectric constant of the dielectric and
the surface optical phonon (SO) energies are obtained from the bulk
longitudinal optical phonon (LO) phonons as $\hbar\omega_{SO}=\hbar\omega_{LO}(\tfrac{1+1/\epsilon_{low}}{1+1/\epsilon_{high}})^{1/2}$.
In this work, we consider only the two SO modes with the strongest coupling strength, denoted as $\omega_{1}$ and $\omega_2$.
$F_{j}$ is the electron coupling parameter with the SO modes. For internal phonon modes,
the energies of OP ($\Gamma$ and K) and their deformation potential $D_{op}$ used in this work are summarized,
and the sound velocity $v_S$ and deformation potential for the AP mode, $D_{ac}$.
See also the respective Appendix for details. \\
 }
\nonumber
\begin{tabular}{ccc}
\hline\hline
\,\,\,\,\,\,\, SPP \,\,\,\,\,\,\,  & \,\,\,\,\,\,\, SiO$_2$\cite{PA10} \,\,\,\,\,\,\, & \,\,\,\,\,\,\, h-BN\cite{BN_par} \,\,\,\,\,\,\, \\
\hline
\multicolumn{1}{c|}{$\epsilon_{low}$} & 3.9 & 5.09\\
\multicolumn{1}{c|}{$\epsilon_{high}$} & 2.4 & 4.1\\
\multicolumn{1}{c|}{$\hbar\omega_1$ (meV)} & 58.9 & 101.7\\
\multicolumn{1}{c|}{$\hbar\omega_2$ (meV)} & 156.4 & 195.7\\
\multicolumn{1}{c|}{$F_{1}^2$ (meV)} & 0.237 & 0.258\\
\multicolumn{1}{c|}{$F_{2}^2$ (meV)} & 1.612 & 0.52\\
\hline\hline
\end{tabular} 
\,\,\,\,\,\,\,\,\,\,\,\,\,\,
\begin{tabular}{ccc}
\hline\hline
\,\,\,\,\,\,\, OP \,\,\,\,\,\,\, & \,\,\,\,\,\,\, $\Gamma$ \,\,\,\,\,\,\, & \,\,\,\,\,\,\, K \,\,\,\,\,\,\, \\
\hline
\multicolumn{1}{c|}{$D_{op}$ (eV\AA$^{\mbox{-}1}$)} & 11 & 16 \\
\multicolumn{1}{c|}{$\hbar\omega_0$ (meV)} & 197  & 157\\
\hline\hline\\
\hline\hline
\multicolumn{3}{c}{AP} \\
\hline
\multicolumn{1}{c|}{$D_{ac}$ (eV)} & \multicolumn{2}{c}{7.1}\\ 
\multicolumn{1}{c|}{$v_S$ (km/s)} & \multicolumn{2}{c}{17}\\
\hline\hline
\end{tabular}
\end{table}

\section{\label{sec:results} Results and discussions}
\subsection{Competing cooling pathways}

\begin{figure}[p]
\centering
\scalebox{0.55}[0.55]{\includegraphics*[viewport=215 205 575 660]{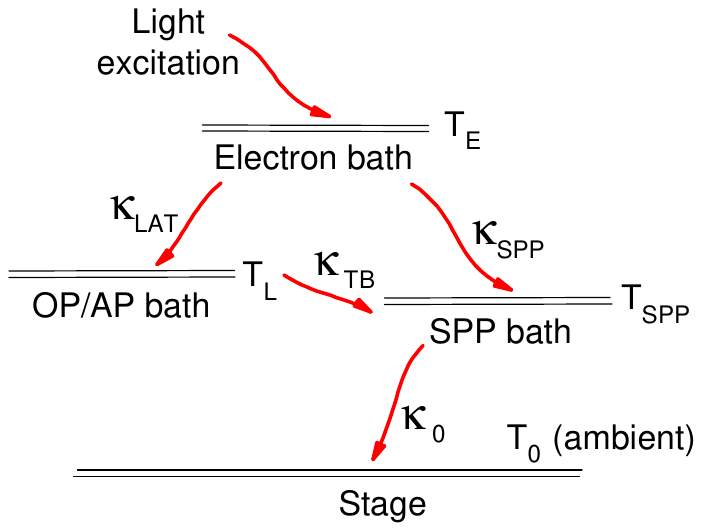}}
\caption{
 (Color online) Cartoon illustrating the typical cooling pathways of hot electrons 
 produced by continuous photoexcitation of graphene with detail descriptions in the main text.
 Heat can also be dissipated through metallic contacts attached to graphene (not shown),
 as discussed in Sec.\,\ref{sec:cooldy}.
 }
\label{fig1}
\end{figure}

We begin with a simple illustration of the possible cooling pathways 
for photoexcited carriers in graphene in Fig.\,\ref{fig1}.
Each thermal bath can be characterized by their respective temperatures $T_{\alpha}$,
and are in general different from the ambient temperature $T_0$.
The heat exchange between these thermal baths can 
be described by the thermal conductivity, $\kappa$,
defined as the ratio between the power exchange
per unit temperature difference i.e. $\delta P/\delta T$.
In general, the different phonon baths can each establish a
different temperature upon interactions with the electrons (see also discussion in Sec.\,\ref{sec:cooldy}).
In this work, we shall assume a common temperature for 
all these phonon baths, denoted simply as the lattice temperature $T_L$.\\

For a typical SiO$_2$ substrate thickness of $h=50-300\,$nm,
$\kappa_0=\kappa/h$ varies in a range $\approx 5\,$MW/Km$^2 - 10\,$MW/Km$^2$,
where SiO$_2$ film thermal conductivity is $\kappa=0.5-1.4\,$W/mK\cite{YNKT02}. 
The interface thermal conductance of graphene on SiO$_2$ substrate
has been measured using various experimental techniques\cite{Balan11,Chen_APL_2009,Freitag_NL_2009,Heinz_APL_2010,Pop_NL_2010},
with values ranging from $\approx 25\,$MW/Km$^2 - 180\,$MW/Km$^2$. 
On the theory front, several approaches have been employed 
to estimate this interface thermal conductance\cite{Pop_PRB_2010,PU10,Persson_2010,Volokitin_2011,Rotkin_unpub},
which varies from $\approx 1\,$MW/Km$^2 - 100\,$MW/Km$^2$.
As illustrated in Fig.\,\ref{fig1}, energy transfered from 
electrons to the internal phonon baths  
is conducted to the underlying substrate 
through a phonon-limited $\kappa_{TB}$. $\kappa_{TB}$ between carbon surface 
and SiO$_2$ substrate has been estimated from molecular dynamics
and is $\approx 60\,$MW/Km$^2$\cite{Pop_PRB_2010}, and can depend
also on the surface roughness. 
Alternatively, energy can be transfered directly to the 
substrate via the SPP phonons, i.e. $\kappa_{SPP}$,
and can depend sensitively on doping.
For an undoped graphene, $\kappa_{SPP}$ is on the order of $1\,$MW/Km$^2$
while $\kappa_{LAT}$ is even smaller, 
as we will see later in the discussion.
In this section, we discuss how these cooling pathways
depend on the various experimental conditions.\\

\begin{figure}[t]
\centering
\scalebox{1.0}[1.0]{\includegraphics*[viewport=183 36 600 580]{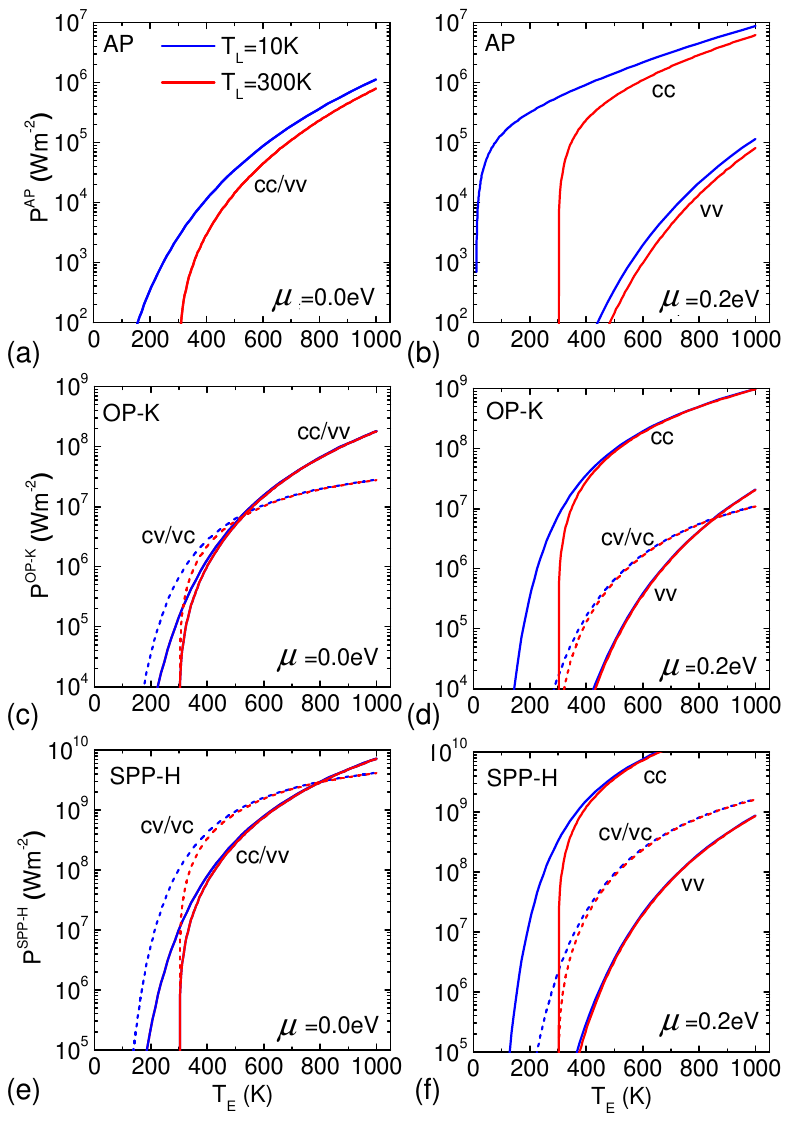}}
\caption{ (Color online) \textbf{(a-b)} Electron cooling power due to acoustic phonons bath, ${\cal P}^{AP}$, as function
of electron temperature $T_{E}$ for different lattice temperature $T_{L}$ calculated
for neutral and doped graphene respectively.
Intraband ($cc$, $vv$) and interband ($cv$, $vc$) processes are indicated.
\textbf{(c-d)} Similar, except for $K$ optical phonons modes, ${\cal P}^{OP,K}$ and
\textbf{(e-f)} for high energy \emph{unscreened} SPP mode, ${\cal P}^{SPP,H}$.
$\Gamma$ optical and the low energy SPP phonons show similar
characteristic (not shown).
}
\label{fig2}
\end{figure}

Detailed balance condition of in- and out-scattering processes requires that ${\cal P}^{\alpha}$
vanishes under equilibrium condition i.e. $\delta T\equiv T_E - T_L=0$.
In the theory, this is ensured by the composite Fermi-Boson distribution function ${\cal F}(T_E,T_L)$.
The energy exchange efficiency with these various phonon baths depends upon, among other factors,
the doping and electronic/lattice temperatures. Using the models described in Sec.\,\ref{sec:theory},
we calculate ${\cal P}^{\alpha}(T_E)$ due to the various phonon baths for intrinsic/doped graphene
under cold/hot (defined at $T_L=10,$ $300\,$K respectively) lattice temperature as shown in Fig.\,\ref{fig2}.\\

First, we discuss results on intrinsic graphene, see Fig.\,\ref{fig2}(a,c,e), which can be understood 
on the basis of scattering phase space arguments.
For cold neutral graphene, Pauli blocking limits the electronic transitions involved to mainly interband processes.
Hence, under near equilibrium condition i.e. $\delta T$ being small,
we observed that the cooling power is mainly dominated by interband processes by optical and SPP modes.
Increasing the electronic temperature alleviates Pauli blocking,
and allows for intraband processes to take place.
As $\delta T$ increases further, we observe that intraband cooling begins to dominate over the interband counterpart.
The efficiency of energy exchange can be explained by
the electron-phonon occupation number, quantified by
the composite distribution function ${\cal F}(T_E,T_L)$ defined in Eq.\,\ref{comF}. For inelastic processes,
one can show that ${\cal F}(T_E,T_L)$ is independent of $T_L$ when $\delta T\rightarrow\infty$.
This is as reflected in Fig.\,\ref{fig2} for ${\cal P}^{OP,K}$ and ${\cal P}^{SPP,H}$.
On the other hand, for quasi-elastic acoustic phonons, the cooling power is proportional $\delta T/T_E$ instead.\\

The results on doped graphene are shown in Fig.\,\ref{fig2}(b,d,f).
Contrary to the intrinsic case, Pauli blocking promotes intraband electronic
transitions over interband processes in doped graphene.
In addition, ${\cal P}^{\alpha}_{cc}\neq {\cal P}^{\alpha}_{vv}$,
with larger cooling power for the majority carriers.
At moderate doping of $\mu = 0.2\,$eV, their cooling power
differs by more than an order of magnitude.
The reduced electron-hole symmetry upon doping also leads to smaller
interband cooling power.
Quasi-elasticity of acoustic phonon scattering results in
a phase space restriction in the scattering, with a Bloch-Gr$\ddot{u}$neisen temperature
determined by the doping\cite{HS08,EK10}, i.e. $T_{BG}=2\hbar v_{S}k_F /k_B$, in contrast to normal metals. 
This increase in phase space in conjunction with Pauli blocking
greatly enhances the cooling power due to AP over the optical phonon baths.
In fact, for moderate $T_E\lesssim 100\,$K, ${\cal P}^{AP}$ dominates over all other mechanisms
for cold graphene.\\

\begin{figure}[p]
\centering
\scalebox{1.0}[1.0]{\includegraphics*[viewport=206 170 545 496]{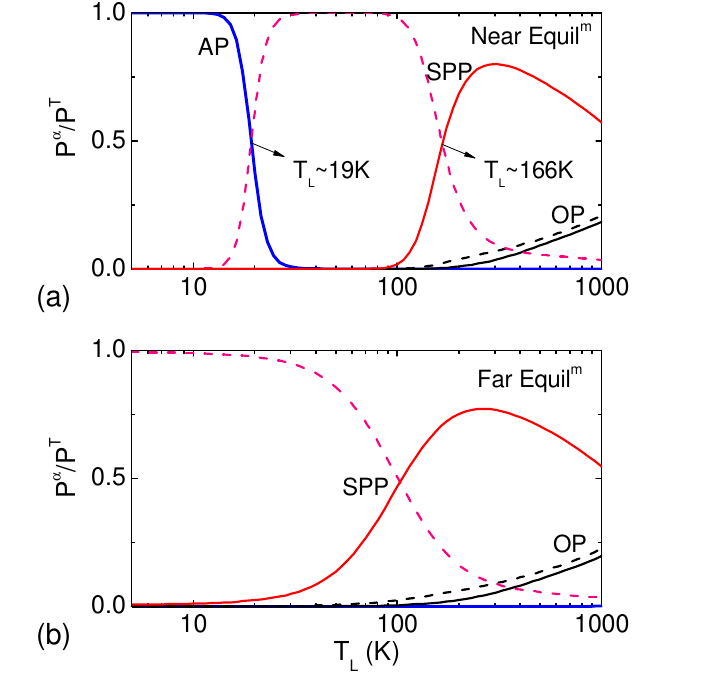}}
\caption{ (Color online) $\bold{(a)}$
Fractional cooling power ${\cal P}^{\alpha}/{\cal P}^{T}$ where ${\cal P}^{T}=\sum _{\alpha}{\cal P}^{\alpha}$,
calculated at ``near equilibrium"' condition of $T_E-T_L=10\,$K for neutral graphene,
including 2D screening $\epsilon_{2D}(q)$ described in text.
For SPP and OP, the dashed (solid) line
represents the low (high) energy mode. $\bold{(b)}$ Same as (a),
except calculated for ``far from equilibrium"' condition of $T_E-T_L=100\,$K.
}
\label{fig3}
\end{figure}

The lattice temperature, $T_L$, also plays an important role in the
competing cooling pathways. Fig.\,\ref{fig3} compares the fractional cooling powers ${\cal P}^\alpha$/${\cal P}^{T}$ for intrinsic graphene,
where ${\cal P}^{T}=\sum_{\alpha}{\cal P}^\alpha$.
To obtain a quantitative estimate, we include in-plane screening of the SPP scattering potential in graphene.
The screening is incorporated through a standard procedure\cite{FG97}
$\left|M_{\bold{k},\bold{k}'}\right|$$\rightarrow$$\left|M_{\bold{k},\bold{k}'}\right|/\epsilon_{2D}(\bold{q},\omega)$.
For simplicity, we employed the static screening dielectric function $\epsilon_{2D}(\bold{q},0)$, which
in the long-wavelength limit assumes a simple form\cite{Ando06b}  $\epsilon_{2D}(\bold{q},0)$$\approx$$ 1+q_s/q$,
where $q_s$=$ e^2/2\epsilon_{0}\kappa \int \tfrac{\partial f}{\partial \epsilon} {\cal D}(\epsilon)d\epsilon$
and ${\cal D}$ is graphene density-of-states. $f$ is the Fermi distribution function
and is a function of the electronic temperature.\\

We analyze the results in two non-equilibrium temperature limits, namely
``near equilibrium" ($T_E-T_L=10\,$K) and ``far from equilibrium" ($T_E-T_L=100\,$K) conditions.
Fig.\,\ref{fig3}a considers the condition of ``near equilibrium".
At low $T_L$, AP dominates cooling. Increasing $T_L$ populates the low-energy SPP
mode, which begins to overtake the cooling power at a temperature of $\sim 20\,$K.
This transition temperature increases with doping e.g. is $\sim 50\,$K at a doping of $0.1\,$eV.
The low energy SPP mode is overtaken by its high-energy mode at $\sim 170\,$K.
A downturn in the high-energy SPP cooling power is observed, due to larger screening
at higher temperatures. Eventually, the optical phonons overtake the SPP for temperatures larger than $1000\,$K.
Fig.\,\ref{fig3}b considers the condition of ``far from equilibrium".
In this case, the SPP dominates the cooling power for all $T_L$, except
at temperature $>$$1000\,$K where optical phonons begin to overtake it.

\subsection{\label{sec:cooldy} Cooling dynamics}

We are interested in the role played by these various phonon baths on the cooling dynamics of photoexcited carriers,
more specifically, the temporal evolution of $T_E$. 
The acoustic and optical phonon baths can each establish a
different temperature upon interactions with the electrons, but processes such as
anharmonic phonon-phonon scattering serve to thermalize them on a picosecond time scale\cite{KPSFW05,BLMM07,Heinz_PRL_2008,YSMC09}.
In this work, we shall assume a common lattice temperature $T_L$,
but acknowledge that in experiments with ultrafast pump-probe, this will not hold true. 
On the other hand, under continuous light excitation, coupling to the heat sink
via the supporting substrate substantially cools the lattice
temperature to within a few degrees Kelvin of the ambient temperature $T_0$
under usual photoexcitation conditions\cite{FLXA12}.
Typically, $T_E-T_L$$\gg 1 K$ under low/moderate excitation power levels used in our studies.
In this regard, the relatively small differences among the various phonon baths can be safely ignored.\\

Hot carriers dynamics can be probed
through optical measurements\cite{SHAH92,SWDL08,MSCRS08,KPSFW05}.
Following a pulsed light excitation, the temporal evolution of carrier relaxation, quantified by its electronic
temperature $T_E$, can be measured using differential transmission spectroscopy.
The electron dynamics are usually described by $\Delta T_E\propto \mbox{exp}(-t/\tau_E)$,
and can be estimated with\cite{SRL11},
\begin{eqnarray}
\tau_E = {\cal C} \left(\frac{d{\cal P}^T}{dT_E}\right)^{-1}
\label{tauE}
\end{eqnarray}
where ${\cal C}=d{\cal E}/dT_E$ is the electron specific heat
and ${\cal E}$ is the energy density of graphene.
In this work, ${\cal C}$ is computed numerically.  However, we note that
for $T_E\ll \mu/k_B$, ${\cal C}$ increases linearly with $T_E$, i.e.
${\cal C}\approx \tfrac{2\pi^2}{3}{\cal D}(\mu)k_B^2 T_E$.
Having computed the total cooling power ${\cal P}^T$ in Sec.\,\ref{sec:results},
$\tau_E$ can be obtained directly from Eq.\,\ref{tauE}.
\\

\begin{figure}[p]
\centering
\scalebox{1.3}[1.3]{\includegraphics*[viewport=215 230 545 460]{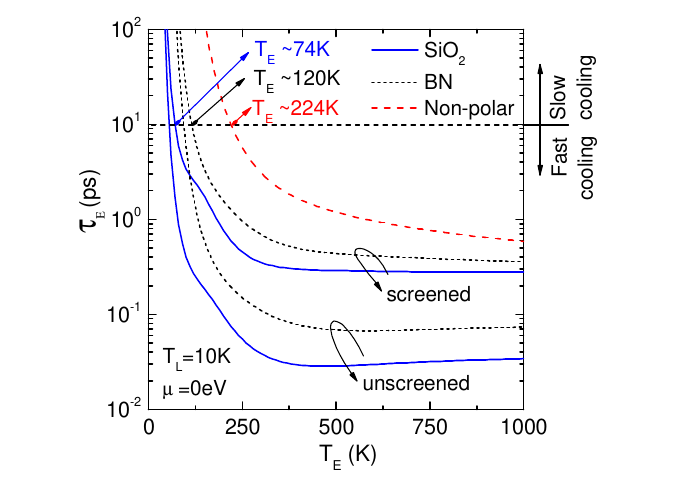}}
\caption{ (Color online)
Carrier's cooling time $\tau_E$ after photoexcitation plotted as function of
$T_E$ calculated for cold neutral graphene. Various substrates
are considered, namely SiO$_2$, BN and non-polar, calculated for
screened and unscreened SPP scattering potentials.
Dotted line distinguishing fast/slow cooling is rather arbitrary, 
and serve only as guide to the eye.
}
\label{fig4}
\end{figure}

Fig.\,\ref{fig4} plots the cooling time, $\tau_E$, for neutral graphene at $T_L=10\,$K.
It is calculated for common substrates such as SiO$_2$, BN and non-polar substrate
such as diamond. At very hot electron temperatures, i.e. $T_E$$>$$500\,$K,
$\tau_E$ is given by a relatively constant sub-picosecond
cooling time.
This is in agreement with experiments\cite{SAJ12}.
The constancy of $\tau_E$ suggests the $\mbox{exp}(-t/\tau_E)$ decay
characteristics typical during the initial fast cooling process.
As $T_E$ cools down, the cooling bottleneck due to AP begins to set in,
leading to much slower cooling times.
The transition temperature into this slow cooling regime varies
with the choice of substrate as indicated in Fig.\,\ref{fig4}.
This transition temperature is dictated by the lowest
frequency SPP mode of the substrate.
Unscreened results, which overestimate the SPP cooling power, yield much
shorter cooling lifetimes than experimentally reported\cite{SAJ12}.
We also note that inclusion of disorder assisted cooling\cite{SRL11},
might enhance the decay rate, especially in the slow cooling regime.
\\

The optoelectronic response in graphene, photovoltaic\cite{XMLGA09,LBW08,PAR09}
or thermoelectric\cite{GSM11,XGA09,SAJ12}, is also a measure of the
energy transport of these hot carriers.
These experiments are usually performed  under a continuous light
illumination of an electrostatic junction.
Their relative contribution depends on the electrostatic junction characteristics,
doping, and even extrinsic factors such as electron-hole puddles\cite{SRML11}.
Nevertheless, at steady state, the photovoltaic current is proportional to the
photo-generated excess carrier density, $\delta n$, via $e\delta n\mu_n \xi$
where $\mu_n$ is the carrier mobility and $\xi$ the local electric field. 
The thermoelectric response, on the other hand, is proportional to the
local elevated temperature, $\delta T$=$T_E-T_L$, via $\sigma(S_1-S_2)\delta T$
where $S_{1,2}$ is the Seebeck coefficient of the two junction
and $\sigma$ is the device conductivity.
Here, we discuss estimates of $\delta n$ and $\delta T$.\\

\begin{figure}[p]
\centering
\scalebox{1.1}[1.1]{\includegraphics*[viewport=215 100 545 555]{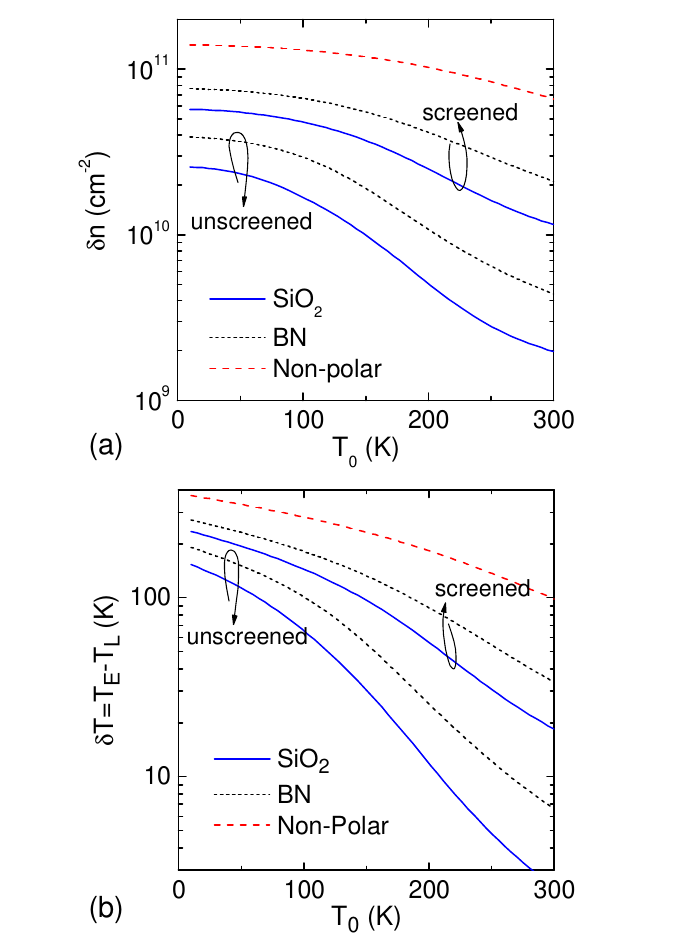}}
\caption{
$\bold{(a)}$  (Color online) Steady-state excess carrier density, $\delta n$, upon continuous photoexcitation
as function of the ambient temperature for neutral graphene. Various substrates
are considered, namely SiO$_2$, BN and non-polar, calculated for
screened and unscreened SPP scattering potentials.
$\bold{(b)}$ Elevated temperatures, $T_E-T_L$, calculated for same conditions in (a).
All calculations assumed ${\cal P}^0 = 1\times 10^7\,$W/m$^{2}$ and $\kappa_0 = 10\,$MW/Km$^{2}$.
}
\label{fig5}
\end{figure}

Under steady state condition,
\begin{eqnarray}
{\cal P}^{0}=\sum_{\alpha} {\cal P}^{\alpha}+{\cal P}^{M}
\label{heatcons}
\end{eqnarray}
where ${\cal P}^{0}$ is the laser power absorbed by graphene, and
${\cal P}^{M}$ is the heat dissipation via the metallic contacts, if any.
In the absence of contacts, all heat dissipation is
via the supporting substrate. At steady state,
${\cal P}^{AP}+{\cal P}^{OP}+{\cal P}^{SPP}\approx (T_L-T_0)\kappa_{0}$.
Eq.\,\ref{heatcons} is then solved self-consistently
in conjunction with charge conservation i.e. $\delta n=n_{e}(T_E,\mu)-n_{e}(T_0,\mu_0)$
=$n_{h}(T_E,\mu)-n_{h}(T_0,\mu_0)$, arriving at steady state values for $T_E$ and $\mu$.
The photoexcited carrier density, $\delta n$, and the elevated temperature,
$\delta T$ are plotted in Fig.\,\ref{fig5}(a-b) respectively,
assuming typical experimental values of ${\cal P}^0 = 1\times 10^7\,$W/m$^{2}$ and 
$\kappa_0 = 10\,$MW/Km$^{2}$.
Both  $\delta n$ and $\delta T$  decrease with increasing ambient temperature $T_0$,
due to more efficient cooling as phonon occupation increases.
For SiO$_2$, $\delta n\approx 10^{10}\,$cm$^{-2}$ and $\delta T\approx 10\,$K
under room temperature condition, of the same order typically seen in measurement\cite{FLXA12}.
Increasing doping increases the scattering phase space for intraband processes and lead to
more efficient electronic cooling into the phonon baths, as shown 
in Fig.\,\ref{fig2}. As a result, both $\delta n$ and $\delta T$
decrease with increasing doping.\\

We also estimate the heat dissipation via contacts phenomenologically with
${\cal P}^M$$\approx$ $\int(f-f_0)(E-\mu){\cal D}/\tau_M dE$ where $f_0$ is
the distribution function before light excitation.
First, we consider the simple ballistic limit where $\tau_M$ is just the
device lifetime given by $L/v_F$, where $L$ is the length of the device and $v_F\approx 10^6$ m/s is the Fermi velocity.
Here, we assume a typical $\tau_M=1\,$ps.  We found that including ${\cal P}^M$ only
leads to few-fold decrease in the quantitative results presented in Fig.\,\ref{fig5} (not shown).
In the realistic case where the carrier transport is in the diffusive
dominated regime, ${\cal P}^M$ would be even smaller, by a factor
of $\lambda/L$, where $\lambda$ is the carrier's mean free path.\\

\begin{figure}[p]
\centering
\scalebox{1.3}[1.3]{\includegraphics*[viewport=215 205 545 460]{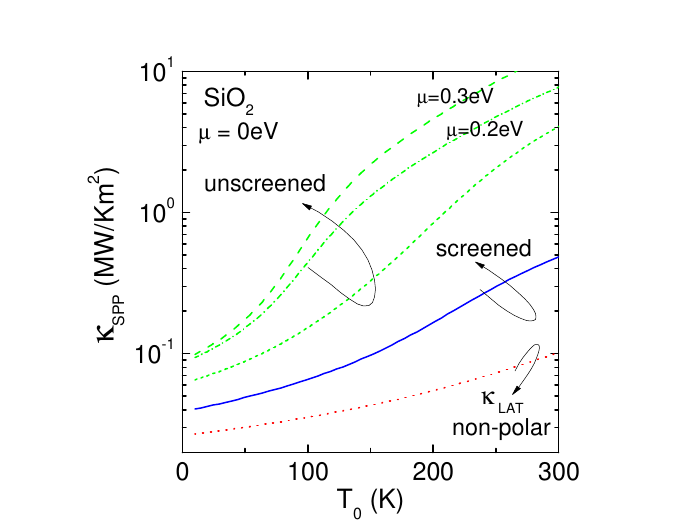}}
\caption{
 (Color online) Out-of-plane thermal conductance, $\kappa_{SPP}$, defined
 as $\kappa_{SPP}={\cal P^{SPP}}/\delta T$ calculated for SiO$_2$ for 
 different conditions such as (i) screened and unscreened SPP scattering potentials
 and (ii) different doping $\mu$ (all curves are for zero doping
 unless stated otherwise). $\kappa_{LAT}$ for undoped graphene on a 
 non-polar substrate is plotted as reference.
}
\label{fig6}
\end{figure}

Fig.\,\ref{fig5} also suggests an order of magnitude enhancement in
the optoelectronic response of graphene, by suppressing the SPP heat dissipation
through a non-polar substrate, such as diamond-like carbon\cite{WLBJ11}, or by suspending graphene.
In fact, the amount of heat transfer to the substrate via the electron coupling with the SPP
can be quantified by an out-of-plane thermal conductance $\kappa_{SPP}$, defined as $\kappa_{SPP}={\cal P}^{SPP}/\delta T$.
This quantity sets the lower limit on the interfacial thermal conductance and it is plotted in Fig.\,\ref{fig6}.
At room temperature, $\kappa_{SPP}\approx 1$MW/Km$^2$ for undoped graphene,
and can increases with doping to order of $10$MW/Km$^2$, see also Ref.\,\cite{Rotkin_unpub}. 
For typical photocurrent experiments, $L$ is typically $\gg \lambda$, and transport is in the diffusive regime.
Here, the in-plane electronic thermal conductivity, $\kappa_e$, can be estimated from the Wiedemann Franz relation.
We found that our estimated value of $\kappa_{SPP}$ is significantly 
larger than $\kappa_e/L^2$ for typical experimental situations.
This suggests that out-of-plane heat dissipation via SPP dominates over the in-plane electronic heat conduction.
The former leads to an increased temperature of the graphene lattice. 
This result reconciles with recent experiment\cite{FLXA12}, 
which reveals significant lattice heating upon laser excitation.
From the experiment\cite{FLXA12}, we can estimate an 
out-of-plane thermal conductance of $\kappa_{exp}=P^0/\delta T\approx 10\,$MW/Km$^2$.
This value is consistent with our estimated $\kappa_{SPP}$.
In fact, $\kappa_{LAT}$ alone is orders of magnitude smaller 
than the experiment as shown in Fig.\,\ref{fig6}.

\section{\label{conclude} Conclusions}


Our results point to the limiting role played by remote substrate phonons in the energy relaxation of hot
photocarriers. In particular, we have shown that the steady state photoresponse in graphene is controlled by the inelastic scattering. The photovoltaic current is proportional to the photo-generated excess carrier density. The thermoelectric contribution, on the other hand, is proportional to the elevated electron temperature. Our results show that irrespective of the mechanism, the SPP phonons limits the overall strength of the photocurrent response on polar substrates.  We predict that a choice of a non-polar substrate will lead to an order of magnitude enhancement in graphene photoresponse.  Therefore, substrate engineering presents a promising route to efficient optoelectronic devices driven by hot carrier dynamics. 
\\

\emph{Acknowledgements:}
TL acknowledges use of a computing cluster
provided by Network for Computational Nanotechnology,
partial funding from INDEX-NRI and in part by the NSF under Grant No. NSF PHY05-51164 (KITP).
We thank F. Xia, H. Yan, F. Guinea, E. Hwang and X. Xu for useful discussions.

\appendix

\section{Acoustic phonon}

We consider first the energy exchange with the acoustic phonon (AP) bath.
The total matrix element for electron-acoustic phonon scattering
due to the two acoustic phonon modes, i.e. $\Gamma_{LA}$ and $\Gamma_{TA}$,
is given by\cite{HS08,PA10},
\begin{eqnarray}
\left|M^{AP}_{\bold{k},\bold{k}'}\right|^2=\frac{ D_{ac}^2 \hbar q}{2\rho_{m}v_S}
\end{eqnarray}
where $D_{ac}$ is the acoustic deformation potential, taken to be $7.1\,$eV in our calulations,  which is very 
similar to the recent ab-initio calculations of $6.8\,$eV\cite{Kaasbjerg_2012}.
We note that the electron-phonon matrix element for these two acoustic modes have
different angular dependencies with transition matrix elements\cite{PSZR80,Kaasbjerg_2012},
which became negated after summing them\cite{PSZR80}.
$v_S$ is graphene effective sound velocity defined as\cite{PA10} $2v_{S}^{-2}=v_{LA}^{-2}+v_{TA}^{-2}$,
where $v_S=17\,$km/s, $v_{LA}=24\,$km/s and $v_{TA}=14\,$km/s.
$\rho_{m}$ is graphene mass density taken to be $7.6\times 10^{-7}\,$kg/m$^2$.
The acoustic phonon is then described by an effective Debye linear dispersion $\omega_{q}=v_{S}q$.
Since $v_S \ll v_F$, $\hbar \omega_q$ is typically much smaller than other energy scale in the problem.
The acoustic phonon scattering is thus approximated to be elastic\cite{BM09} i.e. $k'\approx k$.\\

The cooling power can then be written as,
\begin{eqnarray}
\nonumber
{\cal P}^{AP}_{cc}&=&\frac{g_s g_v }{(2\pi)^2\hbar } \int_{0}^{\infty}k dk \int_{0}^{\infty}k' dk' \int_{0}^{2\pi} d\theta
\frac{D_{ac}^{2}\hbar q}{2\rho_{m}v_S}
\delta(k'-k-\omega_{q}v_{F}^{-1})(k'-k) {\cal F}(k,k')\\
&\approx&\frac{g_s g_v }{(2\pi)^2\hbar } \frac{D_{ac}^{2}\hbar }{2\rho_{m}v_S}\int_{0}^{\infty}  \int_{0}^{2\pi} k^2 q^2 \frac{v_S}{v_F} {\cal F}(k,k) dk d\theta
\end{eqnarray}
Under the assumption $\hbar\omega_{q}\ll T_E, T_L$, we have
\begin{eqnarray}
{\cal F}(k,k)\approx (1-f_{k})f_{k}\frac{T_E-T_L}{T_E}
\end{eqnarray}
Making use of the relations $q^2\approx 2k^2(1-\mbox{cos}\theta)$,
we then obtain a simplified form for the cooling power,
\begin{eqnarray}
{\cal P}^{AP}_{cc}\approx \frac{g_s g_v D_{ac}^{2}}{2\pi \rho_{m}v_S}
\frac{T_E-T_L}{T_E}
\int_{0}^{\infty}k^4 (1-f_{k})f_{k} dk
\end{eqnarray}
Contributions from interband processes, ${\cal P}^{AP}_{cv,vc}$,
are forbidden due to energy-momentum conservations.

\section{Optical phonons}

Next, we discuss energy exchanges with high energy dispersionless optical phonon (OP) modes
i.e. $\Gamma_{LO}$, $\Gamma_{TO}$ and $\mbox{K}_{TO}$.
We consider first the electron-phonon coupling of long-wavelength optical phonon modes,
$\Gamma_{LO}$ and $\Gamma_{TO}$. Their sum is expressed as\cite{LPM05,ANDO06},
\begin{eqnarray}
\left|M^{OP,\Gamma}_{\bold{k},\bold{k}'}\right|^2=\frac{ \hbar D_{OP,\Gamma}^2  }{2\rho_{m}\omega_{o}}
\end{eqnarray}
where $D_{OP,\Gamma}=3\sqrt{2}g/2\approx 11\,$eV$\AA^{-1}$ is the
optical-phonon deformation potential with a coupling constant of $g=5.3\,$eV$\AA^{-1}$,
and $\hbar\omega_{o}=197\,$meV.
We note that the electron-phonon matrix element for these two optical modes have
different angular dependencies with transition matrix elements\cite{LPM05,ANDO06}
i.e. $ 1\pm ss'\mbox{cos}(\theta_{\bold{k}}-\theta_{\bold{k}'})$ where $s=\pm 1$ denotes conduction/valence bands,
which again became negated after summing them.\\

We consider first the intraband cooling power, written as,
\begin{eqnarray}
\nonumber
{\cal P}^{OP,\Gamma}_{cc}&=&\frac{g_s g_v }{(2\pi)^2\hbar } \int_{0}^{\infty}k dk \int_{0}^{\infty}k' dk' \int_{0}^{2\pi} d\theta
\frac{D_{OP,\Gamma}^{2}\hbar }{2\rho_{m}\omega_0}
\delta(k'-k-\omega_{0}v_{F}^{-1})(k'-k) {\cal F}(k,k')\\
\nonumber
&=&\frac{g_s g_v D_{OP,\Gamma}^{2}}{4\pi \rho_{m}v_{F}}\int_{0}^{\infty}k
(k+\omega_{0}v_{F}^{-1}){\cal F}(k,k+\omega_{0}v_{F}^{-1})dk\\
&=&\frac{g_s g_v D_{OP,\Gamma}^{2}}{4\pi \rho_{m}v_{F}}N_{\omega_0}
\left[\mbox{exp}(\frac{\hbar\omega_0}{k_{B}T_{L}})-\mbox{exp}(\frac{\hbar\omega_0}{k_{B}T_{E}})\right]
\int_{0}^{\infty}k(k+\omega_{0}v_{F}^{-1})(1-f_k)f_{k+\omega_{0}v_{F}^{-1}}dk
\end{eqnarray}
In similar fashion, the interband cooling power is written as,
\begin{eqnarray}
{\cal P}^{OP,\Gamma}_{cv}&=&\frac{g_s g_v D_{OP,\Gamma}^{2}}{4\pi \rho_{m}v_{F}}N_{\mbox{-}\omega_0}
\left[\mbox{exp}(\frac{-\hbar\omega_0}{k_{B}T_{L}})-\mbox{exp}(\frac{-\hbar\omega_0}{k_{B}T_{E}})\right]
\int_{0}^{\infty}k(\omega_{0}v_{F}^{-1}-k)H_v[\omega_{0}v_{F}^{-1}-k](1-f_k)f_{k-\omega_{0}v_{F}^{-1}}dk
\end{eqnarray}
where $H_v$ is the Heaviside function.\\

For zone edge phonon modes, only the transverse $\mbox{K}_{TO}$
contributes to carrier scattering, and the matrix element is\cite{LPM05},
\begin{eqnarray}
\left|M^{OP,K}_{\bold{k},\bold{k}'}\right|^2=\frac{ \hbar D_{OP,K}^2  }{2\rho_{m}\omega_{o}}\frac{1-ss'\mbox{cos}\theta}{2}
\end{eqnarray}
where $D_{OP,K}=3g \approx 16\,$eV$\AA^{-1}$ and $\hbar\omega_0=157\,$meV.
The cooling power is similar to the $\Gamma$ phonons case,
except a factor of $\tfrac{1}{2}$ smaller due to the angular dependence.

\section{Surface polar phonons}

The surface polar phonons coupling is given by\cite{FG08,FNC01,PA10},
\begin{eqnarray}
\left|M^{SPP}_{\bold{k},\bold{k}'}\right|^2=\frac{\pi e^2}{\epsilon_0} F_{j}^{2}\frac{\mbox{exp}(-2qz_0)}{q}\frac{1+ss'\mbox{cos}\theta}{2}
\end{eqnarray}
where $\epsilon_0$ is the free space permittivity and $z_0$ is the separation between
graphene and the substrate. The magnitude of the polarization field
is given by the Frohlich coupling parameter, $F_{j}^{2}$.
In common SiO$_2$ dielectrics, there are two dominant surface optical phonon modes
having energies $\hbar \omega_{1}=58.9\,$meV and $\hbar \omega_{2}=156.4\,$meV,
with Frohlich coupling $F_{1}^{2}=0.237\,$meV and $F_{2}^{2}=1.612\,$meV respectively\cite{PA10}.\\

We consider first the intraband cooling power, written as,
\begin{eqnarray}
\nonumber
{\cal P}^{SPP}_{cc}&=&\frac{g_s g_v }{(2\pi)^2\hbar } \int_{0}^{\infty}k dk \int_{0}^{\infty}k' dk' \int_{0}^{2\pi} d\theta
\frac{\pi e^2}{\epsilon_0} F_{j}^{2}\frac{\mbox{exp}(-2qz_0)}{q}\frac{1+ss'\mbox{cos}\theta}{2}
\delta(k'-k-\omega_{j}v_{F}^{-1})(k'-k) {\cal F}(k,k')\\
&=&\frac{g_s g_v }{(2\pi)^2\hbar }\frac{\omega_j}{v_F}\frac{\pi e^2}{\epsilon_0} F_{j}^{2} \int_{0}^{\infty}  k(k+\omega_{j}v_{F}^{-1}) {\cal F}(k,k+\omega_{j}v_{F}^{-1}) \int_{0}^{2\pi}
\frac{\mbox{exp}(-2qz_0)}{q}\frac{1+\mbox{cos}\theta}{2}   d\theta  dk
\end{eqnarray}
The phonon momentum $q$ has the constraint $q^2=k'^2+k^2-2k'k\mbox{cos}\theta$.
Under typical conditions, the factor $\mbox{exp}(-2qz_0)\approx 1$.
Linearizing $\mbox{exp}(-2qz_0)$, the intraband cooling power then becomes,
\begin{eqnarray}
{\cal P}^{SPP}_{cc}&=&\frac{g_s g_v \omega_j\pi e^2 F_{j}^{2}}{(2\pi)^2\hbar v_F \epsilon_0} N_{\omega_j}
\left[\mbox{exp}(\frac{\hbar\omega_j}{k_{B}T_{L}})-\mbox{exp}(\frac{\hbar\omega_j}{k_{B}T_{E}})\right] \int_{0}^{\infty}  kk' (1-f_k)f_{k'} \Theta(k,k') \mbox{exp}\left(\frac{-z_0}{\Theta(k,k')}\right) dk
\end{eqnarray}
where $k'\equiv k+\omega_{0}v_{F}^{-1}$ and
\begin{eqnarray}
\nonumber
\Theta(k,k')&\equiv&\int_{0}^{2\pi}
\frac{1+\mbox{cos}\theta}{2\sqrt{k'^2+k^2-2k'k\mbox{cos}\theta}}   d\theta \\
&=& \frac{k+k'}{kk'}
\left[I_{K}\left(\frac{2\sqrt{kk'}}{k+k'}\right)-I_{E}\left(\frac{2\sqrt{kk'}}{k+k'}\right)\right]
\end{eqnarray}
where $I_{K,E}$ are the complete elliptic integrals of first and second kind.
In similar fashion, the interband cooling power is written as,
\begin{eqnarray}
\nonumber
{\cal P}^{SPP}_{cv}&=&\frac{g_s g_v \omega_j\pi e^2 F_{j}^{2}}{(2\pi)^2\hbar v_F \epsilon_0} N_{\mbox{-}\omega_j}
\left[\mbox{exp}(\frac{-\hbar\omega_j}{k_{B}T_{L}})-\mbox{exp}(\frac{-\hbar\omega_j}{k_{B}T_{E}})\right]\times\\
&& \int_{0}^{\infty}  kk' H_v[k'](1-f_k)f_{-k'}
\Theta(k,-k')\mbox{exp}\left(\frac{-z_0}{\Theta(k,-k')}\right) dk
\end{eqnarray}
where $k'\equiv \omega_{0}v_{F}^{-1}-k$.

\newpage

\end{document}